\documentclass[aps, reprint, prl, twocolumn, superscriptaddress,amsmath,amssymb,noeprint]{revtex4-2}
\usepackage{times}
\usepackage{amsmath}
\usepackage{amssymb}
\usepackage{xfrac}
\usepackage{dsfont}
\usepackage{graphicx}
\usepackage{float}
\usepackage[table]{xcolor}
\usepackage{braket}
\usepackage{bbold}
\usepackage[normalem]{ulem}
\usepackage{bm}
\usepackage{bbm}
\usepackage{color}
\usepackage[dvipsnames]{xcolor}

\usepackage{pdfpages}
\usepackage{pgffor}
\usepackage{multirow}
\usepackage{booktabs}
\makeatletter
\AtBeginDocument{\let\LS@rot\@undefined}
\makeatother

\usepackage[colorlinks=true, urlcolor=blue, linkcolor=blue, citecolor=blue]{hyperref}
\usepackage{siunitx}

\usepackage[T1]{fontenc} 

\newcommand{\berkeleyphy}{Department of Physics, University of California, Berkeley, California 94720}
\newcommand{\CIQC}{Challenge Institute for Quantum Computation, University of California, Berkeley, California 94720}
\newcommand{\LBL}{Materials Sciences Division, Lawrence Berkeley National Laboratory, Berkeley, California 94720}
\newcommand{\columbia}{Department of Physics, Columbia University, New York, NY 10027}

\begin{document}

\title{Rapid Cavity-Based Mid-Circuit Measurement and Feedforward in a Neutral Atom Array}

\author{Tsai-Chen Lee}
\affiliation{\berkeleyphy}
\affiliation{\CIQC}

\author{Jacquelyn Ho}
\affiliation{\berkeleyphy}
\affiliation{\CIQC}

\author{Yue-Hui Lu}
\affiliation{\berkeleyphy}
\affiliation{\CIQC}

\author{Tai Xiang}
\affiliation{\berkeleyphy}
\affiliation{\CIQC}

\author{Nathaniel B. Vilas}
\affiliation{\berkeleyphy}
\affiliation{\CIQC}

\author{Zhenjie Yan}
\affiliation{\berkeleyphy}
\affiliation{\columbia}


\author{Dan M. Stamper-Kurn}
\email[]{dmsk@berkeley.edu}
\affiliation{\berkeleyphy}
\affiliation{\CIQC}
\affiliation{\LBL}

\begin{abstract}

Measuring part of a quantum system in the midst of its evolution and acting on the result in real time is essential for numerous quantum information protocols. Neutral-atom arrays are a leading platform for quantum information processing, but their mid-circuit measurement-and-feedforward cycle times have remained slow, typically exceeding 1 ms. Here we demonstrate fast mid-circuit measurement and real-time feedforward in an array of atomic qubits coupled to a high-finesse optical cavity. Local light shifts tune individual data qubits out of resonance with the cavity, shielding their coherence, while a near-resonant probe drives a selected qubit whose emission is collected with Purcell enhancement.  Mid-circuit measurements of four qubits with sub-percent infidelity reduce the coherence of a fifth unmeasured data qubit by less than 2\%.  We implement real-time feedforward to correct measurement-induced phase shifts and to realize an adaptive circuit for optimal quantum-state discrimination and conditional state preparation. Our approach reduces the measurement-and-feedforward cycle time to below 100~$\mu$s and establishes optical cavities as a route to fast control of neutral-atom quantum systems. 

\end{abstract}

\maketitle

The state of an evolving quantum system becomes nondeterministic owing both to inherent quantum uncertainty and to errors incurred from imprecise operations or environmental decoherence.  Mid-circuit measurement, wherein information is extracted from a portion of a quantum system, and feedforward control conditioned on such information allows one to steer quantum systems toward more certain outcomes and functions.  Mid-circuit measurement and feedforward is central to many quantum information applications, including error correction \cite{Shor1996QEC,Dennis2002ToplogicalQEC,Steane1996QEC,bluvstein2025fault,Katabarwa2024Early,krinner2022realizing}, measurement-based computation~\cite{raussendorf2001one-way}, adaptive circuits \cite{fossfeig2023experimental,griffiths1996semiclassical,baumer2024qft,antonine2025mcmasprimitive}, optimized measurement  \cite{peres1991optimal,conlon2023approaching,Zhou2025Three-copy}, preparation and stabilization of entangled states~\cite{lu2022measurement,Smith2024Constant,feldmeier2026digital,fossfeig2023experimental}, and studies of measurement-induced phase transitions~\cite{li2018quantumzeno,skinner2019measurement-induced,chan2019unitary-projective}.

Neutral atom arrays have emerged as a leading platform for quantum science~\cite{kaufman2021quantum}.  Mid-circuit measurement \cite{dordevic2021entanglement,mcmemma,mcmsaffman,mcmbernien,norcia2023midcircuit,mcmjeff,mcmkaufman,bluvstein2024logical,bluvstein2025fault,wang2026multi,lib2026velocityenabledquantumcomputingneutral} and adaptive feedforward protocols ~\cite{mcmbernien,huie2023repetitive,mcmkaufman,bluvstein2024logical,evered2025probing} have been demonstrated in these systems. However, measurement-and-feedforward cycles in neutral atom systems remain limited to $\gtrsim 1$~ms, more than three orders of magnitude slower than typical coherent quantum operations. This speed limit is set by the long times needed to transport atoms to dedicated measurement zones and to collect photons fluoresced into free space.

In this work, we speed up mid-circuit measurement and feedforward control in a neutral-atom qubit array by combining two techniques. To eliminate the time needed for atom transport, we use site-selective \emph{in-situ} light-induced energy shifts to separate the optical resonances of unmeasured data qubits from those of unshifted measured qubits, protecting them from mid-circuit optical measurement errors \cite{norcia2023midcircuit,mcmvladan}.  To reduce the time needed for photodetection, we couple the array strongly to an optical cavity,  enabling rapid Purcell-enhanced collection of photons emitted by measured qubits that are exposed to a site-selective near-resonant optical drive \cite{Bochmann2010Detection,mcmemma,wang2024ultrafasthighfidelitystatereadout,mcmvladan}. 
Sequential mid-circuit measurements of four atomic qubits with sub-percent infidelity are found to reduce the coherence of a fifth unmeasured data qubit by less than 2\%.
Furthermore, we demonstrate two applications of rapid feedforward control conditioned on mid-circuit measurement outcomes: adaptive correction of coherent phase shifts on data qubits that are contingent on the states of measured qubits, and an adaptive quantum circuit for quantum-state discrimination and coherent state preparation.  Altogether, we reduce measurement-and-feedforward cycle times to as low as 45 $\mu$s, improving on current free-space atom array approaches by an order of magnitude.

\begin{figure}
    \centering
    \includegraphics{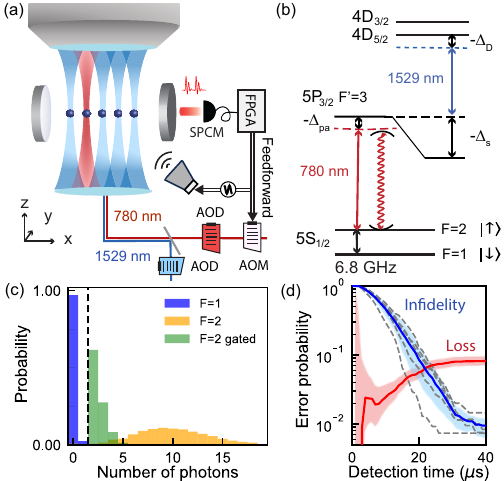}
    \caption{Cavity detection of a 5-qubit atom array. (a) Experimental setup. 
    AODs direct probe and shielding light onto each atom and AOMs control the light intensities.  A single-photon counting module (SPCM) detects cavity emission.  An FPGA processes the SPCM signal for adaptively gated measurement and feedforward control of a microwave driving field. (b) Probe light is detuned from the  $F\!=\!2$ to $F'\!=\!3$ transition and resonant with the cavity. Shielding light shifts the average frequency of the $F^\prime =3$ levels, decoupling the unmeasured qubits from near-resonant light. (c) Histograms of cavity-emission photon counts in 40~$\mu$s of cavity detection for atoms prepared in the $F=1$ manifold (blue), and in the $|F=2,m_F=0\rangle$ state without (orange) and with (green) adaptive gating for all the atoms in the array. In this data set, probe light is detuned by $\Delta_\mathrm{pa}= -2\pi \times 20$~MHz and undetected atoms are shielded by $\Delta_\mathrm{s}=-2\pi\times700$ MHz. (d) Detection infidelity (blue) and atom loss (red) as a function of maximum detection time with adaptive gating for atoms prepared in $|F=2,m_F=0\rangle$.  Infidelity is quantified by limiting the photon detection window, and loss by post-selecting events according to the detection time. Gray dashed lines denote detection infidelities for each of the 5 atoms. Shading indicates uncertainty (68\% confidence interval) in the array-averaged values.
    }
    \label{fig:scheme}
\end{figure}

In our apparatus \cite{cavitymicroscope,mcmemma}, we load a one-dimensional tweezer array containing five $^{87}$Rb atoms into a near-concentric Fabry--Perot optical cavity whose TEM$_{00}$ mode resonance is near the D$_2$ transition of $^{87}$Rb at a wavelength of $\lambda=780$~nm (Fig.\ \ref{fig:scheme}(a)). The atoms are placed along the cavity axis with $19.5\,\mu$m ($25 \lambda$) spacing, on antinodes of the cavity mode. The maximal single-atom cooperativity is $C=g_0^2/2\kappa\gamma=3.0$, with $\{g_0,\kappa,\gamma\}=2\pi\times\{3.1,0.53,3.0\}$~MHz. Here, $2g_0$ is the single-photon Rabi frequency for an atom at a cavity antinode driven on the $\ket{F=2,m_F=2}\rightarrow \ket{F'=3,m_F'=3}$ cycling transition, and $\kappa$ and $\gamma \equiv \Gamma/2$ are the half-linewidths of the cavity and atomic resonances, respectively.

Each ground-state atom encodes one qubit in a hyperfine-state basis.  To prepare atoms in the $|F=2, m_F =0\rangle \equiv \ket{\uparrow}$ hyperfine state, we apply a 3 G magnetic field along an axis ($z$) that is perpendicular to the cavity axis ($x$) and illuminate the atoms with $\pi$-polarized light resonant with the $F=2 \rightarrow F^\prime =2$ transition and additional repump light resonant with the $F=1 \rightarrow F^\prime = 2$ transition, both on the $\mathrm{D}_2$ line.  Coherent single-qubit gates are effected by exposing the array to microwave pulses that drive atoms to the $\left|F=1, m_F = 0 \right\rangle \equiv \ket{\downarrow}$ state.

We perform mid-circuit measurement by illuminating the atom array with two colors of light; see Figs.\ \ref{fig:scheme}(a) and (b).  A 780-nm-wavelength probe beam, slightly detuned (by $\Delta_\mathrm{pa}$) from the cycling transition and resonant with the cavity, is focused onto a single measurement qubit.  A measured atom initially in the $F=2$ hyperfine manifold scatters photons brightly into the cavity while an atom initially in the $F=1$ hyperfine manifold is dark.  By counting probe photons emitted from the cavity, we discriminate between the logical states of the measured qubit.

Simultaneously, shielding beams, at a wavelength of 1529 nm and detuned by $\Delta_\mathrm{D}$ from the $5\mathrm{P}_{3/2}\rightarrow4\mathrm{D}_{5/2}$ atomic transition, are focused onto each of the remaining data qubits.  The ac Stark effect of these shielding beams shifts the energies of the $5 \mathrm{P}_{3/2}$ states, tuning the $\mathrm{D}_2$ optical transitions of shielded atoms away from those of measured atoms.  One can quantify this shielding by an average frequency shift $\Delta_\mathrm{s}$ on the $\ket{5 \mathrm{P}_{3/2}, F^\prime = 3}$ states, which increases in magnitude with the shielding beam optical intensity and decreases with $|\Delta_\mathrm{D}|$; 
see the Supplemental Material~\cite{supplemental}.
The offset optical response of the shielded atoms reduces their sensitivity to cavity photons and protects their coherence during mid-circuit measurement.  The shielding light has little effect on the ground-state coherence of shielded atoms~\cite{supplemental}.  Both sets of beams are steered through acousto-optic deflectors (AODs), allowing for site-selective
readout of the array.

We report first on the detection infidelity for measurement qubits. Here, to reduce the impact of imperfect $m_F$ state optical pumping, we replace the state preparation in the $\ket{\downarrow}$ state with optical depumping into the $F=1$ manifold using a depump beam
that addresses the $F=2\rightarrow F'=2$ transition. Since our measurement is not $m_F$-state-discriminating, the bright state infidelity is unaffected. After state preparation, the array is sequentially measured using  $40~\mu \mathrm{s}$ detection windows, with 20~$\mu \mathrm{s}$ switching intervals between atoms. Our circularly polarized probe beam  pumps bright-state atoms into the stretched state.  This $m_F$ state pumping reduces the probability of optical depumping of bright atoms into the $F=1$ manifold. As shown in Fig.\ \ref{fig:scheme}(c), the histograms of cavity emission photocounts during each detection window discriminate strongly between bright and dark atomic hyperfine states.  Setting a minimum threshold of two photons to detect a bright-state atom, we achieve per-atom detection infidelities as low as $0.9^{+0.4}_{-0.3}\% $ for the bright state, and $0.2^{+0.4}_{-0.2}\%$ for the dark state, averaged over the array. 
The bright state readout infidelity is limited by recoil-induced mechanical effects and early-time depumping, which stops the atom from fluorescing into the cavity before two photons are collected; see Supplemental Material \cite{supplemental}.

While longer probe times ideally allow for larger photon counts and greater state discrimination (Fig. \ref{fig:scheme}(d)), they also raise the probability of losing the measured atom from its tweezer owing to recoil forces and heating, and increase the disturbance of data qubits owing to residual interactions with cavity photons. 
To reduce these deleterious effects while maintaining low measurement infidelity, we adopt an adaptive gating method using a field programmable gate array (FPGA) that shuts off 
the probe light, through an acousto-optical modulator (AOM) switch, after two photons are detected~\cite{mcmvladan,Chow2023high-fidelity}. This gating decreases the atom loss rate from 95\% to an average of 8\%, and reduces the mean bright-state detection time to $9.5~\mu$s. 
 \begin{figure}
    \includegraphics{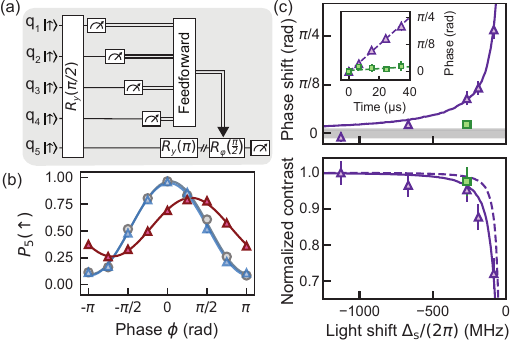}

    \caption{Mid-circuit measurement and phase shift feedforward correction. (a) Experimental sequence for the Ramsey measurement with echo. 
    (b) Probability $P_{5}(\uparrow)$ of atom $q_5$ being measured in the $\ket{\uparrow}$ state after mid-circuit measurements of $q_1$-$q_4$ with shielded detunings of $\Delta_\mathrm{s} = -2\pi\times 1130$ MHz (blue triangles) and $\Delta_\mathrm{s} = -2\pi \times 80$ MHz (red triangles). Reference data taken without mid-circuit measurement are shown as gray circles.
    Solid curves are sinusoidal fits. 
    (c) Ramsey phase shift (top) and normalized contrast (bottom) vs.\ $\Delta_\mathrm{s}$
    without (purple triangles) and with (green squares) feedforward correction. The gray shaded bar represents the uncertainty in the phase of the reference data. Solid curves show theoretical calculations that account for 
    photon scattering and stochastic variations in the accumulated phase shift, without phase shift correction. The dashed curve shows the fundamental limit to contrast due to interaction with cavity photons~\cite{supplemental}. All data are taken at $\Delta_\mathrm{D} = -2\pi \times 12$~GHz, with the exception of the points at $\Delta_\mathrm{s} = - 2\pi \times 1130$~MHz, which use $\Delta_\mathrm{D} = -2\pi \times 6$~GHz for stronger shielding. 
    Inset: Phase shift vs. total mid-circuit detection time with (green squares) and without (purple triangles) phase shift feedforward correction, at $\Delta_\mathrm{s} = -2\pi\times 270$~MHz. 
    Data points are binned into 10~$\mu$s time intervals. Error bars denote one standard error of the mean.
    The measurements in this figure are made with a probe detuning $\Delta_\text{pa} = -2\pi \times 10$~MHz and a maximum detection time of 30~$\mu$s, with 
    similar infidelity as in Fig~\ref{fig:scheme}. We do not observe a systematic dependence of infidelity on the shielding strength. The largest variation (an increase of infidelity by $3\%$) occurs at $\Delta_\mathrm{s} = -2\pi\times 80$~MHz for the final measured atom, which may experience more depumping during the preceding measurements.
    }
    \label{fig:mcmcircuit}
\end{figure}

To qualify our detection as a mid-circuit measurement, we confirm that the quantum information in data qubits is conserved during the measurement.  To do this, we embed mid-circuit measurements on four qubits within a simple quantum circuit, comprising single-qubit gates, that realizes a Ramsey pulse sequence on a fifth qubit (Fig.\ \ref{fig:mcmcircuit}(a)).  Specifically, all five atoms in the array are prepared in the $\ket{\uparrow}$ state and then subjected to a global $\pi/2$ rotation in the logical basis.  Four atoms (labeled $q_1$ through $q_4$) are measured sequentially using the local probe, optical shielding, and adaptive gating methods described above.
A spin-echo $\pi$-pulse is applied and then, after a delay, a final $\pi/2$-pulse is applied with variable microwave phase before the state of the fifth qubit ($q_5$) is read out. The spin echo $\pi$-pulse is equally spaced between the two $\pi/2$-pulses to suppress sensitivity to shot-to-shot variations in the energy difference between the logical basis states induced by the tweezer trapping light.

The resulting Ramsey contrast curves for the data qubit with and without the mid-circuit measurements are shown in Fig.~\ref{fig:mcmcircuit}(b).
Without mid-circuit measurements on the first four qubits, we observe a baseline Ramsey contrast of 87(1)\% (gray circles), limited by the fidelity of state preparation into the $\left|\uparrow\right\rangle$ state (89(2)\%).
With measurements on the four qubits and a shifted detuning of $\Delta_\mathrm{s}=-2\pi\times1130$ MHz, the same contrast is observed (blue triangles) with a Ramsey phase shift of $\Delta\phi= 0.03(5)$ rad relative to the baseline data. We define the normalized contrast $\mathcal{V}=1.00(3)$ as the ratio of the Ramsey contrast with mid-circuit measurement to the baseline contrast. 
The state-preserving fidelity \cite{Jozsa1994fidelity} after mid-circuit measurements is
$\mathcal{F} \equiv (\operatorname{Tr} \sqrt{\rho^{1 / 2} \rho_{\mathrm{ref}} \rho^{1 / 2}})^2=\frac{1}{2}+\mathrm{Re}(\rho_{\uparrow\downarrow})=\frac{1}{2}(1+\mathcal{V}\operatorname{cos}\Delta\phi) = 100(2)\%$. Here $\rho$ ($\rho_\mathrm{ref}$) is the density matrix of the data qubit right after (right before) the mid-circuit detection, 
and $\rho_{\uparrow\downarrow}$ is the coherence of the data qubit after mid-circuit readout in the $\{\left |\uparrow\right\rangle,\left|\downarrow\right\rangle\}$ basis, measured by the Ramsey experiments.

While near-unity state-preserving fidelity is obtained when applying shielding beams at the maximum shielding condition, weaker shielding (lower $|\Delta_\mathrm{s}|$) of data qubits during mid-circuit measurements leads to reduced coherence and increased Ramsey phase shifts from fluorescence photons circulating in the cavity~\footnote{Direct illumination by probe light focused onto measured atoms has no discernible effect on other atoms in the array; see Supplemental Material \cite{supplemental}.}. The observed Ramsey phase shift and normalized contrast after mid-circuit measurement are plotted in Fig.~\ref{fig:mcmcircuit}(c)  vs.\ $\Delta_\mathrm{s}$. We understand the reduced normalized contrast as coming from two sources:  Spontaneous scattering of photons circulating in the cavity collapses the data qubit, eliminating coherence. Additionally, the data qubit dephases owing to variations in the phase shift accumulated between basis states depending on the random outcomes of mid-circuit measurements, the variable duration of the adaptively gated probe pulse, position fluctuations of data qubits that change their coupling strength to the cavity field, and quantum-optical fluctuations. The observed phase shifts and contrast reductions agree well with theoretical calculations (solid curves) that include all these effects; see Supplemental Material \cite{supplemental}.  

Now we describe the integration of rapid mid-circuit measurement with feedforward control of an atom array.  We present two examples of such integration.

First, we utilize feedforward control to mitigate cross-talk errors in our mid-circuit measurement.  Grouping the Ramsey curves generated by our measurement circuit by the total detection time, conditioned on whether the measured atom is observed to be in the $\ket{\uparrow}$ state, we find that the average phase shift on the data qubit is proportional to the total time the cavity is filled with probe photons. To accentuate this effect, we operate purposely at weak shielding ($\Delta_\mathrm{s} = -2\pi\times 270$ MHz), and observe a mean Ramsey phase shift per detection time of 
$0.020(2)~\text{rad}/\mu\text{s}$ (Fig.\, \ref{fig:mcmcircuit}(c) inset).
We therefore adopt feedforward control in our Ramsey experiment, using our FPGA device to track the total bright-cavity detection time over the four mid-circuit measurements and to shift the phase of the final $\pi/2$-pulse in our Ramsey circuit according to the calibration obtained above.
After feedforward corrections, the Ramsey phase shift slope is reduced to $0.002(2)~\text{rad}/\mu\text{s}$,
and the state-preserving fidelity is increased from $\mathcal{F} = 96(2)\%$ to  $\mathcal{F} = 99(2)\%$. 

 \begin{figure} 
    \centering
\includegraphics[width=0.9\linewidth]{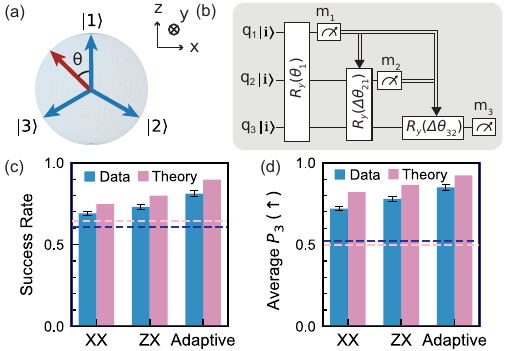}
    \caption{Quantum-state discrimination and coherent state preparation using mid-circuit measurement and feedforward control. (a) Three atoms in an array are prepared identically in one of three possible single-particle states, shown on the Bloch sphere (blue arrows).
    A measurement and feedforward protocol is used to identify the initial state based on measurements on $q_1$ and $q_2$ and to prepare $q_3$ coherently in the $\ket{\uparrow}$ state, as confirmed by a final measurement on $q_3$.
    The measurements are along axes defined by polar angles $\theta_i$ (red arrow) for atoms $q_i$.  (b) Circuit diagram of the experimental protocol.  Global microwave rotations by angles $\Delta \theta_{ij} = \theta_i-\theta_j$ are applied prior to measurements in the $z$ basis; see Tab.\ \ref{tab:1}. 
(c) Experimental (blue) and ideal theoretical (pink) quantum-state discrimination success probability under two fixed-measurement schemes and the optimal adaptive-measurement scheme.  Dashed lines show experimental (blue) and theoretical (pink) baselines where all measurements are made in the $z$ basis. (d) Measured and calculated coherent state preparation success probabilities, determined from measurement outcomes on $q_3$. }
    \label{fig:adaptivemeasurement}
\end{figure}

Second, we use mid-circuit measurement and feedforward control for the following quantum information task: An ensemble of $N$ qubits is prepared randomly in one of several non-orthogonal product states selected from a set $\left\{\ket{j}^{\otimes N}\right\}$.  Qubits are detected sequentially from this ensemble with two different objectives: discriminating which product state was prepared, and coherently preparing an unmeasured qubit from the ensemble in a fiducial state (specifically $\ket{\uparrow}$).  While neither objective can be realized with perfect success, the minimum-error protocol for both state discrimination and conditional coherent state preparation with local measurements requires sequential mid-circuit measurement and feedforward control~\cite{peres1991optimal, chitambar2013revisiting}. 

In the experiment, we identically prepare three atoms (qubits $q_1$-$q_3$) in a state chosen
from the set
$\{ \left | 1\right\rangle = \left|\uparrow\right\rangle$,
$\left|2\right\rangle = \cos(\frac{\pi}{3})\left|\uparrow\right\rangle+\sin(\frac{\pi}{3})\left|\downarrow\right\rangle$,
$\left|3\right\rangle = \cos(\frac{\pi}{3})\left|\uparrow\right\rangle-\sin(\frac{\pi}{3})\left|\downarrow\right\rangle \}$, each with equal probability (Fig. \ref{fig:adaptivemeasurement}(a))~\cite{peres1991optimal,weir2018optimal}. 
Next, we sequentially measure the qubits using the circuit shown in Fig.~\ref{fig:adaptivemeasurement}(b).
Each measurement is made by applying a global microwave rotation followed by a local mid-circuit measurement, such that qubit $q_i$ is measured along an axis defined by the polar angle $\theta_i$, with measurement outcome $m_i$.  
The measurement outcomes $m_1$ and $m_2$ are used to determine the state discrimination success probability by comparing the true prepared state to the inferred state~\cite{supplemental}. The final measurement outcome $m_3$ is used to determine the state preparation success probability, $P_{3}(\uparrow)$.
A 5 $\mu$s wait time after each 40 $\mu$s detection window is used to process measurement information and program the ensuing microwave pulse duration; together, these define a measurement-and-feedforward cycle time of 45 $\mu$s.
The time required to implement the entire sequence of Fig.\ \ref{fig:adaptivemeasurement}(b) is $\sim 300~\mathrm{\mu s}$, limited by the long duration of microwave pulses driven at low Rabi frequency.  

Results for the state discrimination and state preparation success probabilities are shown in Fig.~\ref{fig:adaptivemeasurement}(c) for several detection schemes. Specifically, we demonstrate two suboptimal fixed-axis measurement schemes (the ``XX'' and ``ZX'' schemes) that do not make use of feedforward to select the measurement angle $\theta_2$, and compare them to the optimal adaptive-measurement scheme, where conditional feedforward is used to determine
$\theta_2$ based on measurement outcome $m_1$.
For all three schemes, conditional feedforward is used to select the final measurement axis, $\theta_3$, which is chosen to maximize the probability of measuring the qubit in the bright state. In the lab frame, this maximizes coherent state preparation of qubit $q_3$ in the $\ket{\uparrow}$ state given the measurement outcomes $m_1$ and $m_2$. 
The schemes are summarized in Table~\ref{tab:1} and described further in the Supplemental Material \cite{supplemental}.

\begin{table}[t]
\centering
\renewcommand{\arraystretch}{1.15}
\setlength{\tabcolsep}{7.5pt} 
\begin{tabular}{|c| c @{\hspace{0.5\tabcolsep}} c |c c c|}
\hline
\textbf{Scheme } & $m_1$& $m_2$ & $\theta_1$ & $\theta_2$ & $\theta_3$ \\
\hline
\multirow{4}{*}{\textbf{XX}}
& 1& 1 & $\pi/2$ & $\pi/2$ & $0.58\pi$ \\
\cline{2-6}
& 1& 0 & $\pi/2$ & $\pi/2$ & $0$ \\
\cline{2-6}
& 0& 1 & $\pi/2$ & $\pi/2$ & $0$ \\
\cline{2-6}
& 0& 0 & $\pi/2$ & $\pi/2$ & $-0.58\pi$ \\
\hline
\multirow{4}{*}{\textbf{ZX}}
& 1& 1 & $0$ & $\pi/2$ & $0.15\pi$ \\
\cline{2-6}
& 1& 0 & $0$ & $\pi/2$ & $-0.15\pi$ \\
\cline{2-6}
& 0& 1 & $0$ & $\pi/2$ & $0.69\pi$ \\
\cline{2-6}
& 0& 0 & $0$ & $\pi/2$ & $1.31\pi$ \\
\hline
\multirow{4}{*}{\textbf{Adaptive}}
& 1& 1 & $\pi/6$ & $-0.11\pi$ & $0.01\pi$ \\
\cline{2-6}
& 1& 0 & $\pi/6$ & $-0.11\pi$ & $0.68\pi$ \\
\cline{2-6}
& 0& 1 & $\pi/6$ & $0.44\pi$ & $0.66\pi$ \\
\cline{2-6}
& 0& 0 & $\pi/6$ & $0.44\pi$ & $1.32\pi$ \\
\hline
\end{tabular}
\caption{Schemes for quantum state discrimination and coherent state preparation of qubit $q_3$ conditioned on mid-circuit measurement outcomes $m_1$ and $m_2$ for qubits $q_1$ and $q_2$, respectively. Here, $m_i=1$ ($0$) corresponds to a bright (dark) measurement outcome. Spin projections of $q_i$ along an axis with polar angle $\theta_i$ (in the $x$-$z$ pseudo-spin plane) are measured.
The XX scheme is optimal for the case $\theta_1=\theta_2$, while the ZX scheme is optimal for the case where $\theta_2$ cannot be chosen adaptively.
}
\label{tab:1}
\end{table}

The adaptive-measurement scheme, empowered by mid-circuit measurement and feedforward control, outperforms the fixed-measurement schemes in both objectives. Quantum-state discrimination succeeds with a probability of $81(2)\%$ with adaptive measurement, compared with probabilities of $69(1)\%$ and $73(2)\%$ in the fixed-measurement XX and ZX schemes, respectively. Conditional coherent state preparation produces a $\ket{\uparrow}$ measurement result on $q_3$ with probability $85(2)\%$ with adaptive measurement, compared with probabilities of $72(2)\%$ and $78(2)\%$ in the fixed-measurement XX and ZX schemes, respectively. All three of these schemes outperform a baseline approach in which no state rotations are performed, so that all measurements are made in the initial $z$ basis, for which we measure probabilities of $52(1)\%$ and $61(2)\%$ for the state-discrimination and state-preparation success, respectively.  The underperformance of all three feedforward schemes relative to ideal implementation is largely due to imperfect initial state preparation in the $\ket{\uparrow}$ state ($89(2)\%$ fidelity) and residual microwave-pulse errors.

We have demonstrated fast sequential mid-circuit readout and feedforward control of a neutral-atom qubit array coupled to an optical cavity.  We achieve high-fidelity single-qubit measurements with fixed-duration measurement windows of 30 to 40 $\mu$s, while the optical probe duration within each window is reduced to even shorter times by adaptive gating.  Programming times of 5 $\mu$s are used to initiate conditional microwave pulses and of 20 $\mu$s to switch to a new qubit for mid-circuit measurement.  Altogether, we establish the capacity to implement measurement-and-feedforward control, in a variety of potential applications, with cycle times well below 100 $\mu$s.
Various technical modifications of our setup would further reduce the cycle time, such as increasing the fluorescence detection rate using higher-cooperativity cavities, increased photodetection quantum efficiency, and higher-intensity counter-propagating probe beams; reducing the measurement times upon adaptive gating; 
and implementing much faster gates using all-optical means \cite{levine2022dispersive} (rather than the current, much slower, microwave pulses).  

Rapid mid-circuit measurements and feedforward control enable several future advances in quantum information science. In the area of many-body quantum physics, measuring and acting on a system mid-evolution shows an advantage in preparing long-range entangled states beyond constraints from local gates~\cite{fossfeig2023experimental,lu2022measurement}, in steering systems through measurement-induced phase transitions~\cite{li2018quantumzeno,skinner2019measurement-induced}, and in facilitating efficient simulation of complex Hamiltonians~\cite{alan2005simulated,lanyon2010towards}. The cavity interface for atom arrays is naturally suited to quantum networking and gate teleportation~\cite{covey2023quantum,ramette2022any} and quantum key distribution~\cite{Gottesman2003Proof}.  Augmenting the strongly coupled cavity-tweezer platform with coherent Rydberg quantum gates \cite{de2026realization} would lead to Rydberg quantum processors that are rapidly monitored and stabilized through cavity-based measurement.

\begin{acknowledgments}
We acknowledge support from the AFOSR (Grant No.
FA9550-1910328), from ARO through the DURIP program (Grant No. W911NF2310244), from DARPA (Grant No.
W911NF2010090), from the NSF through the Challenge Institute of Quantum Computing (Grant No. OMA-2016245) and the MRI program (Grant No. OMA-2216201), and from the U.S. Department of Energy (Grant No. 7562496). J.H. acknowledges support from the Department of Defense through the National Defense Science and Engineering
Graduate (NDSEG) Fellowship Program. N.B.V. acknowledges support from the Miller Institute for Basic Research in Science.
\end{acknowledgments}
\bibliography{main.bib}

 \pagebreak

 

\section*{Supplementary material}

# Supplemental Material for Rapid Cavity-Based Mid-Circuit Measurement and Feedforward in a Neutral Atom Array

Tsai-Chen Lee,[1,2] Jacquelyn Ho,[1,2] Yue-Hui Lu,[1,2] Tai Xiang,[1,2]
Nathaniel B. Vilas,[1,2] Zhenjie Yan,[1,3] and Dan M. Stamper-Kurn[1,2,4,*]

[1]*Department of Physics, University of California, Berkeley, California 94720*
[2]*Challenge Institute for Quantum Computation, University of California, Berkeley, California 94720*
[3]*Department of Physics, Columbia University, New York, NY 10027*
[4]*Materials Sciences Division, Lawrence Berkeley National Laboratory, Berkeley, California 94720*

## I. SITE-SELECTIVE ILLUMINATION

In our setup, both the fluorescence probe light (780 nm wavelength) and the shielding light (1529 nm wavelength) are imaged into the vacuum chamber and onto the atom array from below ($z$ axis, perpendicular to the cavity axis). Both beams are re-imaged upon exit from the chamber through the high numerical aperture lens that is used to focus optical tweezer light (from above) onto the atoms. The focused probe and shielding beam waists ($1/e^2$ intensity radius) are measured to be 3.3 $\mu$m and 4.6 $\mu$m, respectively, both smaller than the separation between atoms. With the re-imaging system, we observe no leakage of probe light onto data qubits from illumination of a measurement qubit. As an additional check, we post-select outcomes of the mid-circuit measurement Ramsey circuit (described in the main text in Fig. 2) for events when all four measured atoms are dark (state $|\downarrow\rangle$). The phase shift in the Ramsey signal in those data would reflect the exposure of the data qubit to stray probe illumination. Even at low shielding ($\Delta_{\mathrm{s}} = -2\pi \times 80$ MHz) we discern no Ramsey phase shift or contrast reduction in these data compared with those in which no probe-light illumination is used. This observation supports our conclusion that probe light is focused on single atoms with essentially no direct exposure onto neighboring atoms in the array.

The shielding light, which is focused (onto data qubits) to a larger beam waist than the probe light, does overlap somewhat onto neighboring atoms in the array. We quantify this overlap by detecting the fluorescence brightness in the $\mathrm{D}_2$ optical resonance frequency of one atom while shielding light is focused onto the location of a neighboring site of the array. We observe a light shift that is $2 \times 10^{-3}$ as large as the light shift produced at the shielding beam's focus. Even at the strongest shielding, this degree of overlap does not significantly affect the fidelity of mid-circuit measurement.

## II. SHIELDING WITH 1529-NM-WAVELENGTH LIGHT

Shielding light couples the $5\mathrm{P}_{3/2}$ and $4\mathrm{D}_{3/2,5/2}$ states of $^{87}$Rb with detuning $\Delta_{\mathrm{D}}$ relative to the $5\mathrm{P}_{3/2}(F = 3) \rightarrow 4\mathrm{D}_{5/2}(F = 4)$ transition, as shown in Fig. S1(a). Plots of the ac polarizability of the $5\mathrm{P}_{3/2}$ and $5\mathrm{S}_{1/2}$ states at optical frequencies near resonance with these transitions are shown in Fig. S1(b). These show that the scalar polarizability of $5\mathrm{P}_{3/2}$ is $> 10^4$ times larger than that of the $5\mathrm{S}_{1/2}$ state. The differential polarizability between hyperfine qubit states is $10^{-2}$ a.u at $\Delta_{\mathrm{D}} = -2\pi \times 12$ GHz, and shows little variation over different $\Delta_{\mathrm{D}}$.

At the combination of optical intensities and detunings used in the experiment, the 5P and 4D manifolds are coupled sufficiently strongly that the ac polarizability (based on second order perturbation theory) does not accurately describe the effect of the shielding light on the atoms. To account for this, we diagonalize a Hamiltonian describing electric dipole interactions of shielding light with all hyperfine states within the $5\mathrm{P}_{3/2}$, $4\mathrm{D}_{5/2}$, and $4\mathrm{D}_{3/2}$ manifolds. Fig. S1(c,d) shows energy eigenvalues of this Hamiltonian at several shielding beam powers (on a single atom) and detunings. For comparison, calculations using the atomic polarizability are also shown in Fig. S1(c).

To calibrate the intensity of the 1529-nm-wavelength shielding beams in the experiment precisely, we rely on a "magic" detuning at $\Delta_{\mathrm{D}} = 2\pi \times 24$ GHz, at which frequency the differential polarizability between $m_F$ sublevels in the $5\mathrm{P}_{3/2}(F = 3)$ manifold crosses through zero (see Fig. S1(b)). We illuminate atoms with probe light blue detuned above the $5\mathrm{S}_{1/2}(F = 2) \rightarrow 5\mathrm{P}_{3/2}(F = 3)$ resonance frequency while applying shielding light at the magic detuning with variable power. At specific shielding-beam powers, the atomic transition is brought into resonance with the probe, causing atoms to be resonantly heated out of the tweezers. Comparison with the Hamiltonian described above (which uses known $^{87}$Rb atomic parameters) provides a precise calibration of the laser intensity at each atom site. We observe a variation of approximately 10% in the intensity of the shielding beams across the 5-atom array.

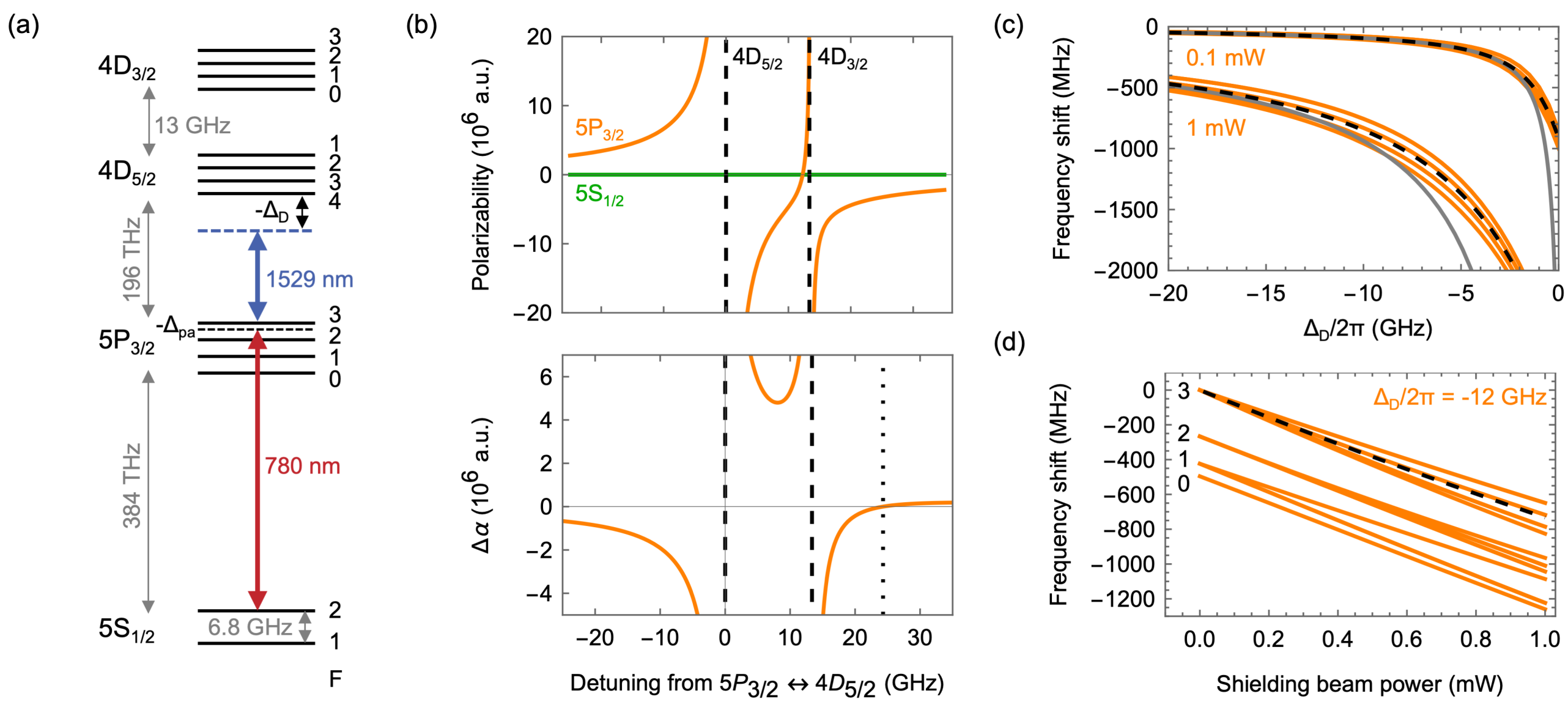


Figure S1. Shielding with 1529-nm-wavelength light. (a) Energy level structure (not to scale). (b) ac polarizabilities of relevant states as a function of $\Delta_\mathrm{D}$. Top: scalar polarizability of $5\mathrm{P}_{3/2}$ and $5\mathrm{S}_{1/2}$. The $5\mathrm{S}_{1/2}$ state has a scalar polarizability of 420 a.u. across the range of detunings in the plot. Bottom: Differential polarizability, $\Delta\alpha$, of magnetic sublevels in the $5\mathrm{P}_{3/2}(F=3)$ manifold, defined here as the difference between the $|F=3,|m_F|=3\rangle$ and the $|F=3,m_F=0\rangle$ state polarizabilities when exposed to linearly polarized light as in the experiment. There is a magic detuning where $\Delta\alpha=0$ at $\Delta_\mathrm{D}\approx 2\pi\times 24$ GHz (vertical dotted line). Polarizablities are plotted in units of 1 a.u. $=4\pi\epsilon_0 a_0^3$. (c) Energy eigenvalues of the $5\mathrm{P}_{3/2}(F=3,m_F)$ states vs. $\Delta_\mathrm{D}$ at shielding powers of 0.1 mW and 1 mW per atom (assuming a 4.6 $\mu$m Gaussian beam waist). For comparison, the expected energies based on the scalar polarizability are shown as solid gray curves. (d) Energy eigenvalues of $5\mathrm{P}_{3/2}$ hyperfine states (labeled on the left of the plot) vs. shielding beam power (with a 4.6 $\mu$m Gaussian waist) at $\Delta_\mathrm{D}=-2\pi\times 12$ GHz. The average $|5\mathrm{P}_{3/2},F=3\rangle$ frequency shifts, $\Delta_\mathrm{s}$, are shown as black dashed curves in (c) and (d).

Away from the magic detuning, the Zeeman states and $F$ manifolds are shifted by different amounts, so there is no unique shielding shift for the transition. Therefore, we quantify the degree of shielding using $\Delta_\mathrm{s}$, which is defined as the average frequency shift of the seven states originating as the $|5\mathrm{P}_{3/2},F=3,m_F\rangle$ states at zero power, relative to the unperturbed $|5\mathrm{P}_{3/2},F=3\rangle$ state energy. $\Delta_\mathrm{s}$ is plotted with dashed curves in Fig. S1(c-d).

## III. ANALYSIS OF DETECTION FIDELITY

The infidelity for bright-state ($F=2$) cavity detection comes from a combination of depumping into $|F=1\rangle$ states, recoil-induced atom loss and heating, and state preparation errors. These effects are explained in more detail below.

Unlike the work in [S1], we do not use cavity detection to discriminate between a tweezer loaded with a dark state atom and a tweezer with no trapped atom. We confirm atom occupation with 200 ms of fluorescence imaging. The photon-count histograms with and without an atom are non-overlapping[S2], allowing for unambiguous occupation discrimination.

During cavity detection, we use circularly polarized probe light that optically pumps the bright state atom into the $|F=2,m_F=2\rangle$ stretched state, where the coupling to the cavity is maximized and the probability of depumping into $F=1$ becomes zero. However, depumping errors can still occur due to imperfect laser polarization, as well as during the first few photon scattering events, before the atoms are fully optically pumped into the stretched state. To understand these optical pumping effects, we modeled the system theoretically with a Rabi frequency of $2\pi\times 15$ MHz for the probe light (defined on the $|F=2,m_F=2\rangle\rightarrow|F'=3,m'_F=3\rangle$ stretched transition) and a probe-atom detuning $\Delta_\mathrm{pa}=-2\pi\times 20$ MHz. With these parameters, we find that the optical pumping process happens in around 5 $\mu$s. With pure $\sigma^+$ polarization, depumping error (i.e., an atom undergoing spontaneous decay to an $|F=1\rangle$ state before two photons are collected) is found to be 0.2% based on Monte-Carlo simulations. With 10% polarization impurity, the infidelity increases to 0.4% after a detection time of 40 $\mu$s.

Probing with a single-sided beam makes the detection more susceptible to mechanical motion and the Doppler shift. Experimentally, we observe that the frequency of the light scattered by the atoms decreases with longer probe times;

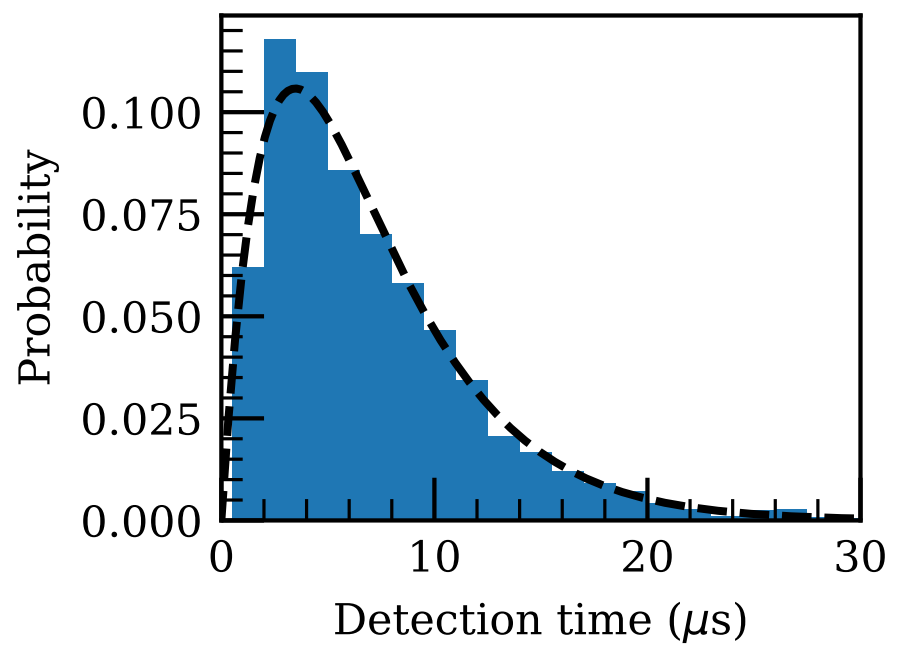


Figure S2. Histogram of times at which the second photon is detected for atoms prepared in $|\uparrow\rangle$. The dashed line is a fit to an Erlang distribution. The mean probe duration is 7 $\mu$s for the detection parameters used in Fig 2 in the main text.

we attribute this to the Doppler shift during the probe process. This effect results in a larger detuning between the scattered light and cavity resonance and leads to a decrease in the late-time signal.

## IV. MEASUREMENT-INDUCED DECOHERENCE ON SHIELDED QUBITS

In our experiment, the phase accumulated by a data qubit during each mid-circuit measurement is affected by four stochastic processes: fluctuations of the atom-cavity coupling due to the finite atomic temperature, the duration of the detection pulse, the state into which the measured atom is projected, and additional quantum fluctuations. We include the first three effects in a Monte Carlo simulation when calculating the theory curves in Fig. 2(c). Quantum-optical fluctuations are accounted for based on an analytic dephasing model.

The thermal motion of trapped atoms along the $x$ axis induces atom-cavity coupling fluctuations since $g(x) = g_0 \cos(kx)$, where $k = 2\pi/(780\text{ nm})$. The temperature of the atoms in the 808-nm-wavelength tweezers is estimated to be 30 $\mu$K from a combination of release-recapture measurements and from the measured Ramsey coherence time, which is limited by thermal motion of the atoms in the tweezers. This position-dependent coupling is included in our model for both the shielded atom and for the atom illuminated by probe light (the latter of which causes variations in the intra-cavity optical field) by sampling over a Gaussian distribution of atomic position.

The probe duration of each detection window $t_{\text{det}}$ is sampled from an Erlang probability distribution $f(t; k_s, \lambda_r)$. The rate parameter $\lambda_r = 0.29$ ct/$\mu$s corresponds to the photon detection rate, and the shape parameter $k_s = 2$ corresponds to 2 photons being received. When the local probe pulse turns on and excites a single atom, it establishes a cavity field if the measured atom is projected into the $|\uparrow\rangle$ state, which occurs with a probability of 50%. After two photons are detected, the probe shuts off. A fit of the Erlang distribution to our experimental data is shown in Fig. S2. Our total cavity photon detection efficiency, including optical losses and detector quantum efficiency, is measured to be 20%. We can therefore infer the average intra-cavity photon number, $\bar{n}_{\text{cav}}$, based on the measured photon detection rate.

We model the phase shift and decoherence on the data qubit for the Ramsey sequence in Fig. 2(a) of the main text using a Monte Carlo simulation. In each realization, the atom-cavity coupling strengths $g$ (and thus the cavity photon number $\bar{n}_{\text{cav}}$), the projected states of the four measurement qubits, and the probe durations ($t_{\text{det}}$) are randomly sampled as described above. Due to large ground state hyperfine splitting, the light circulating in the cavity interacts only with the $|\uparrow\rangle$ state, imposing a coherent phase shift and potentially exciting the atom to emit a photon into free space. In the regime $|\Delta_{\text{s}} - \Delta_{\text{pa}}| \gg g$, it suffices to treat the shielded atoms and the excited single-atom-plus-cavity system separately. Therefore, the phase shift, $\Delta\phi$, and the rate of spontaneous emission to free space, $\Gamma_{\text{sc}}$, during detection time $t_{\text{det}}$ can be calculated by summing over all excited states in the dressed $5\text{P}_{3/2} - 4\text{D}$ manifold:

$$\Delta\phi = \frac{4t_{\rm det}\bar{n}_{\rm cav}|g|^2}{\left|\langle 5{\rm S}_{1/2}||\vec{d}||5{\rm P}_{3/2}\rangle\right|^2} \times \sum_i \frac{\left|\langle F=2, m_F=0|\vec{d}\cdot\hat{\epsilon}_{\rm cav}|i\rangle\right|^2}{\Delta_{\rm pa}-\Delta_{{\rm s},i}} \tag{S1a}$$

$$\Gamma_{\rm sc} = \frac{4\Gamma\bar{n}_{\rm cav}|g|^2}{\left|\langle 5{\rm S}_{1/2}||\vec{d}||5{\rm P}_{3/2}\rangle\right|^2} \times \sum_i \frac{\left|\langle F=2, m_F=0|\vec{d}\cdot\hat{\epsilon}_{\rm cav}|i\rangle\right|^2}{(\Delta_{\rm pa}-\Delta_{{\rm s},i})^2} \tag{S1b}$$

Here, the sum is over all eigenstates $|i\rangle$ of the shielding Hamiltonian described in sec. II (each with eigenenergy $E_i$), $\hat{\epsilon}_{\rm cav}$ is the polarization of the cavity mode, $\Delta_{\rm pa}$ is the cavity detuning from the $5{\rm S}_{1/2}(F=2) \to 5{\rm P}_{3/2}(F=3)$ transition, $\Delta_{{\rm s},i} = [E_i - E_{5{\rm P}_{3/2}}(F=3)]/\hbar$ is the detuning of eigenstate $|i\rangle$ relative to $5{\rm P}_{3/2}(F=3)$, $\vec{d}$ is the electric dipole operator, and $\langle 5{\rm S}_{1/2}||\vec{d}||5{\rm P}_{3/2}\rangle = 5.98$ a.u. is the reduced dipole matrix element of the $D_2$ transition.

For a data qubit initially in the superposition state $|+\rangle = \frac{1}{\sqrt{2}}(|\uparrow\rangle + |\downarrow\rangle)$ exposed to a coherent drive on the bright-to-excited state transition, coherence is destroyed upon a free-space spontaneous photon scattering event. On average, this causes the coherence to decay with time as $\sim e^{-\Gamma_{\rm sc}t/2}$. In a strongly coupled cavity, additional decoherence is expected due to quantum-optical fluctuations in the phase shift induced by cavity photons. In the dispersive regime, the dephasing rate owing to this effect is given by $\Gamma_\phi = 4\theta_s^2\kappa n_{\rm cav}$ [S3], where $\theta_{\rm s} = \Delta\phi/(2\kappa n_{\rm cav}t_{\rm det})$ is the differential phase shift on the qubit states due to interactions with a single photon in the cavity. Therefore, the overall decay of coherence during the mid-circuit detection time is described by $\rho_{\uparrow\downarrow}(t) = \frac{1}{2}e^{-(\Gamma_\phi+\Gamma_{\rm sc}/2)t}$, where $\rho_{\uparrow\downarrow}$ is the coherence of the qubit states.

After averaging the accumulated phase and contrast over thousands of realizations, the simulated normalized contrast is obtained from the absolute value of $\langle 2\rho_{\uparrow\downarrow}e^{i\Delta\phi}\rangle$ for the final state, and the average phase accumulation is given by $\langle\Delta\phi\rangle$. The theory curve is computed at $\Delta_{\rm D} = -2\pi \times 12$ GHz with varied shielding light strength to obtain different $\Delta_{\rm s}$, and matches our data in Fig. 2(c) of the main text. Although one data point in the figure uses $\Delta_{\rm D} = -2\pi \times 6$ GHz, we note that the comparison remains valid because the relevant control parameter is the shifted detuning.

## V. STATE DISCRIMINATION AND PREPARATION WITH MEASUREMENT AND FEED-FORWARD

As described in the main text, we utilize single-atom mid-circuit measurements and feedforward to determine the state of an ensemble of qubits and to prepare a qubit coherently in the fiducial $|\uparrow\rangle$ state. In defining the optimal choices of mid-circuit and final measurement axes, we maximize the probability of coherent preparation of the final (third) qubit. A variety of other metrics have also been used in optimizing state-discrimination problems [S4, S5], including Shannon entropy, minimum error, and others.

We identify the optimal strategy by identifying three polar angles, $\theta_1$, $\theta_2$ and $\tilde{\theta}_2$, with $\theta_1$ defining the axis for measuring the first qubit, and $\theta_2$ ($\tilde{\theta}_2$) defining the axis for measuring the second qubit in the case where the first measurement yields a bright (dark) outcome $m_1 = 1$ ($m_1 = 0$). After two measurements with outcomes $m_1$ and $m_2$, the state of the third qubit is described by the density matrix

$$\rho(m_1, m_2) = \sum_j P(j|m_1m_2)\rho_j, \tag{S2}$$

Here $j \in \{1,2,3\}$ labels the initial states, which are described by density matrices $\rho_j = |j\rangle\langle j|$, where $|j\rangle = \cos(\frac{\phi_j}{2})|\uparrow\rangle + \sin(\frac{\phi_j}{2})|\downarrow\rangle$ and $\phi_j = \{0, \frac{2}{3}\pi, \frac{4}{3}\pi\}$. The posterior probabilities $P(j|m_1m_2)$ are calculated by Bayesian inference and are dependent on $\theta_1, \theta_2, \tilde{\theta}_2$. For each measurement outcome $(m_1, m_2)$, the maximum (initial-state-averaged) probability of measuring qubit $q_3$ in the $|\uparrow\rangle$ state, after a measurement-conditioned coherent rotation, is the maximum eigenvalue of $\rho(m_1, m_2)$, which we label $\lambda_{\rm max}[\rho_{{\rm m}_1,{\rm m}_2}]$. The measurement axis (described by polar angle $\theta_3$) that achieves this maximum is defined by the eigenvector of $\rho(m_1, m_2)$ corresponding to $\lambda_{\rm max}[\rho_{{\rm m}_1,{\rm m}_2}]$.

Numerical optimization was performed to maximize

$$P_3(\uparrow) = \sum_j \sum_{m_1,m_2} P(j, m_1m_2)\lambda_{\rm max}[\rho_{{\rm m}_1,{\rm m}_2}], \tag{S3}$$

with $P(j, m_1m_2)$ being the joint probability of initial state preparation $|j\rangle$ and obtaining measurement outcomes $m_1$ and $m_2$. The angles $\theta_1$, $\theta_2$ and $\tilde{\theta}_2$ defined by this optimization are shown in the main text [S6].

A separate procedure was applied to select angles $\theta_1$, $\theta_2$, and $\tilde{\theta}_2$ to maximize instead the success rate of correctly inferring the state based on mid-circuit measurements of qubits $q_1$ and $q_2$. We found that the angles determined by this separate procedure were all very close (within about 0.02 rad) to those identified above for the state-preparation objective.

* dmsk@berkeley.edu



\end{document}